\documentclass[times,twocolumn,final]{elsarticle}

\usepackage{cag}
\usepackage{framed,multirow}
\usepackage{paralist}

\usepackage{amssymb}

\usepackage{url}

\usepackage{hyperref}

\journal{Computers \& Graphics}

\begin{document}

\verso{Accepted manuscript}

\begin{frontmatter}

\title{Data-driven insight into the puzzle-based cybersecurity training}

\author[1]{Karolína \snm{Dočkalová Burská}\corref{cor1}}
\cortext[cor1]{Corresponding author.}
\ead{burska@mail.muni.cz}
    
\author[2]{Vít \snm{Rusňák}}
\ead{rusnak@ics.muni.cz}
\author[1]{Radek \snm{Ošlejšek}}
\ead{oslejsek@fi.muni.cz}
\address[1]{Faculty of Informatics, Masaryk University, Brno, Czech republic}
\address[2]{Institute of Computer Science, Masaryk University, Brno, Czech republic}

\received{30 July 2021}
\accepted{27 September 2021}

\begin{abstract}

Puzzle-based training is a common type of hands-on activity accompanying formal and informal cybersecurity education, much like programming or other IT skills. However, there is a lack of tools to help the educators with the post-training data analysis.

Through a visualization design study, we designed the Training Analysis Tool that supports learning analysis of a single hands-on session. It allows an in-depth trainee comparison and enables the identification of flaws in puzzle assignments. We also performed a qualitative evaluation with cybersecurity experts and students. The participants apprised the positive influence of the tool on their workflows. Our insights and recommendations could aid the design of future tools supporting educators, even beyond cyber security.
\end{abstract}


\end{frontmatter}


\section{Introduction}\label{s:introduction}

Higher-order thinking has become one of the essential skills for the $21^{st}$ century. The best way to develop and strengthen these abilities is through practical hands-on courses~\cite{medeiros2018,mcmurtrey2008}. One commonly used learning method for training problem-solving or various IT skills (e.g., programming) is puzzle-based learning. Michalewicz et al.~\cite{michalewicz2008} introduced a game-based learning method that uses puzzles as a metaphor for getting students to think about how to frame and solve unstructured problems. In IT education, the puzzle-based learning approach has been prevalent for many years~\cite{yoneyama2008,merrick2010,harms2015}. Even programming courses consist of basic concepts such as \emph{recursion} with assignments like ``Write a program to calculate the factorial of a given number.''

Multiple studies confirmed the usefulness of puzzle-based learning also for cybersecurity education~\cite{gondree2013,hendrix2016,dasgupta2013}. However, while hands-on training produces a tangible output in many learning areas, e.g., a code that can be checked, analyzed, and evaluated, cybersecurity training is process-oriented. Puzzles are tasks like ``search for a vulnerability on server X'' that are difficult to track. Tutors have only a limited view of what trainees are doing in the computer network and how they deal with the task, making the post-training evaluation challenging. 
This paper presents results of cooperation with cybersecurity education experts that led to the design of a visualization tool supporting the follow-up learning analysis of the training sessions.

Regardless of the education subject, tutors make intensive efforts to create, organize, and continually improve these so-called \emph{blended courses}\footnote{Blended courses combine computer-supported learning activities with traditional face-to-face interaction during training sessions.}. Trainees' assessment, which usually follows the training session, is integral to the teaching process. The focus lies on comparing individual trainees and analyzing their progress or discovering weaknesses in the training design.

We contribute to the state of the art of applying visualizations in education practice with:
(a) a user requirement definition on support tools for tutors of the hands-on puzzle-based learning activities (in the cybersecurity education context); (b) design and implementation of the visualization tool for the post hoc analysis of data from the training session; and (c) an evaluation with domain experts resulting in design recommendations for future work.

\section{Related work}
\label{s:related-work}

Assessing the effectiveness of game-based learning poses a significant challenge in the learning analytics research domain. Loh~\cite{loh2012} distinguishes between "assessment \emph{for} learning" and "assessment \emph{of} learning." The former is designed to assess a learner's understanding at the course end. The latter is more helpful to educators because it helps them to improve the learning processes. This paper deals with educators' insight into the learning process. A considerable effort has been made in the past to conceptualize data mining and digital assessment for serious games so that generic learning analytics principles can be researched and applied regardless of the specific game content~\cite{chung2015,alonso2017,owen2020}. Our solution deals with event logs and the score-based assessment that represent broadly accepted types of telemetry and evaluation data for serious games. 

Our work lies at the intersection of education, visualization, and HCI research. According to the classification provided in~\cite{oslejsek2020tvcg}, this paper addresses visual data analysis tasks of organizing participants (referred to as \emph{tutors}). Using information technologies in blended courses enables us to collect metadata produced by learners. Tutors can use them for a post hoc analysis of learners' progression and content revision. Nevertheless, the design and deployment of efficient support tools remain a challenging problem~\cite{Rodriguez-Triana2016}. There are general tools that could be used for specific post-training tasks, e.g., comparing score-based assessment settings via the LineUp application~\cite{gratzl2013}. Our tool aims to reflect the well-defined requirements of training designers and tutors, providing them with a domain-specific comprehensible analytical dashboard.

The purpose of the post-training learning analysis is to understand and optimize learning processes. Previous works~\cite{Matcha2019,Jivet2018,oslejsek2018,deFreitas2017,loh15} address using visual dashboards for learning analysis and confirm the need for insight exceeding simple summative feedback~\cite{macfadyen2010}. Apart from focusing on the learning process, learning analytics in higher education also provide valuable teaching or research resources~\cite{siemens2011}. Analytical tools can support decision-making and improve pedagogical approaches. 

Most of these learning analysis tools focus on the high-level perspective evaluation of students' performance. Existing surveys overview and analyze learning dashboards either for tutors~\cite{Verbert2013a, Verbert2013b, schwendimann2017} or students~\cite{Bodily2017}. Most of them are related to the uptake of massive online open courses. These tools focus on visualizing learning activity, tracking specific learning goals, and providing a high-level perspective on learners' progress. Moodleboard~\cite{Sebastien2019} is a decision support tool for pedagogical engineers and administrators providing both course statistics and detection of flaws or misuses for an open-source learning management system Moodle.
LISSA~\cite{Charleer2018} aims at improving student-advisor dialogue during face-to-face consultations. The tool provides an overview of study progress or peer comparison among multiple students. SAM~\cite{govaerts10} is a general-purpose web-based environment visualizing learners' activities, improving awareness, and supporting self-reflection. Such high-level tools represent domain-independent systems to gather, process, and report the collected and derived data while overlooking disciplinary knowledge practices.

\begin{figure}[!htbp]
  \centering
  \includegraphics[width=.48\textwidth]{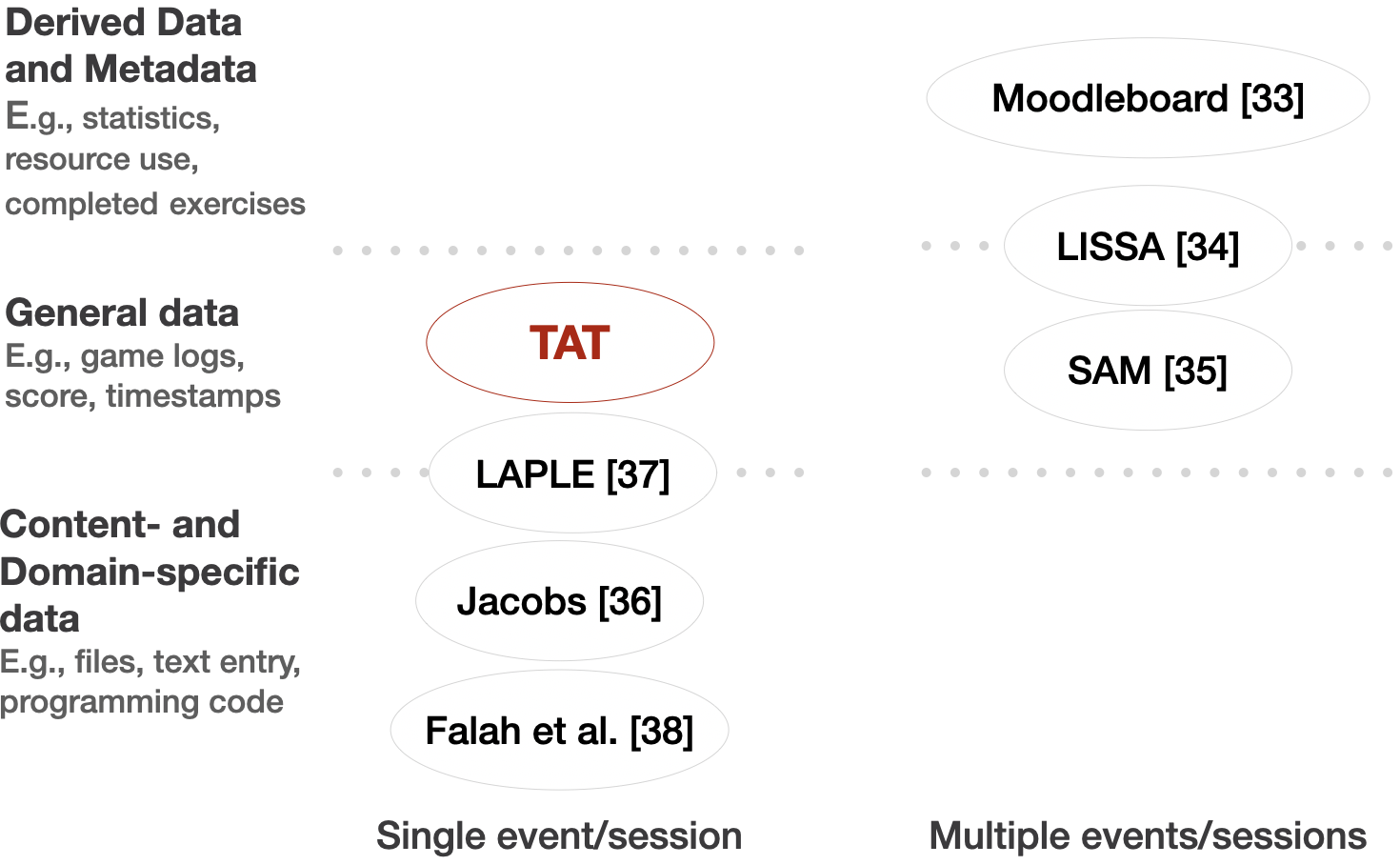}
  \caption{Categorization of learning analytics tools based on their focus (on single or multiple sessions) and the input data types (from domain-specific to derived meta-data). TAT position is highlighted.}
  \label{fig:data-categories}
\end{figure}

In contrast, tools for lower-level data analysis from practical courses often require considering insight from domain experts because the input data driving the analytical tools are domain-specific. Examples can be found for math~\cite{jacobs05}, where the system tackles the understanding of selected math functions, programming tools~\cite{Fu2017} that utilize compilation processes and software quality metrics for assessment, or penetration testing~\cite{falah2017} based on knowledge graphs. \autoref{fig:data-categories} categorizes these tools in two axes: x-axis -- single or multiple training sessions; y-axis -- data specificity, i.e., from the domain-specific data to derived data and metadata.

We propose the \emph{Training Analysis Tool} (TAT) -- a dashboard-like tool for tutors providing data-driven insight into a training session through several linked visualizations. The TAT supports tutors in low-level learning analytics tasks such as inspection and comparison of trainees or identifying training design flaws based on the data from single training sessions.

\section{Background}
\label{s:background}

The puzzle-based learning in the cybersecurity domain is primarily represented by \emph{Capture the Flag} (CTF) games~\cite{werther2011,davis2014,svabensky2018}. CTF training scenarios serve as puzzle-based templates structuring the content into levels focused on solving cybersecurity tasks, e.g., scan the network, identify a server, find the server vulnerability, exploit it, and gain the root privileges. 
CTF games can be organized in diverse ways. Very popular are unsupervised online games when a trainee can access the game or interrupt it anytime. Tutored (or supervised) training sessions for small groups are often practiced in a formal cybersecurity education or professional training. The supervised training sessions share the principles of blended courses popular in primary and secondary education.

CTF games contain a short background story, task assignments, their evaluation, hints, and solutions for each level. A typical scenario consists of up to ten levels. Finding a level solution is necessary to proceed to the next one. Training scenarios use multiple gamification characteristics such as scoring, level-based approach, or scoreboards. Trainees are penalized when taking hints or solutions and reach score points for successful solutions.

Hands-on cybersecurity training is often organized in so-called cyber ranges. The \emph{KYPO Cyber Range Platform}\footnote{\url{https://kypo.cz}} (hereafter referred to as \emph{KYPO CRP})that we use for development and evaluation is a cloud-based environment providing features for the virtualization of computer systems and networks~\cite{celeda2015}. 
It serves as a platform for practical training of various cybersecurity skills in university courses as well as for the training of practitioners from institutions outside. The \emph{KYPO Cyber Range} allows us to create so-called \emph{sandboxes} -- isolated computer networks consisting of multiple virtual machines for several dozens of trainees (the exact number depends on the cloud capacity and resource requirements). The web portal provides a user interface for the management of sandboxes, users, \emph{training scenarios}, and organizing training sessions.

A typical training session is organized for 15--20 participants in the IT classroom. Trainees log in to the web portal and launch a training scenario consisting of a sequence of cybersecurity puzzles. Trainees solve the puzzles individually in their private sandboxes without affecting others' work. A successful solution of the puzzle yields a short string (called \emph{flag}). Entering the flag in the web portal opens the next level. Trainees who are struggling can use hints specific for each level. When helpless, they can see the correct solution (a list of steps leading to the flag). Time for solving all the levels is usually limited to the class length (one or two hours). Tutors walk around and help trainees either on request or when they realize that someone significantly lacks behind (typically by quick peek on their displays or asking them directly). In the end, the scoreboard shows individual scores, and tutors hold a short debriefing to present correct solutions.

\begin{figure}[!htbp]
  \centering
  \includegraphics[width=.48\textwidth]{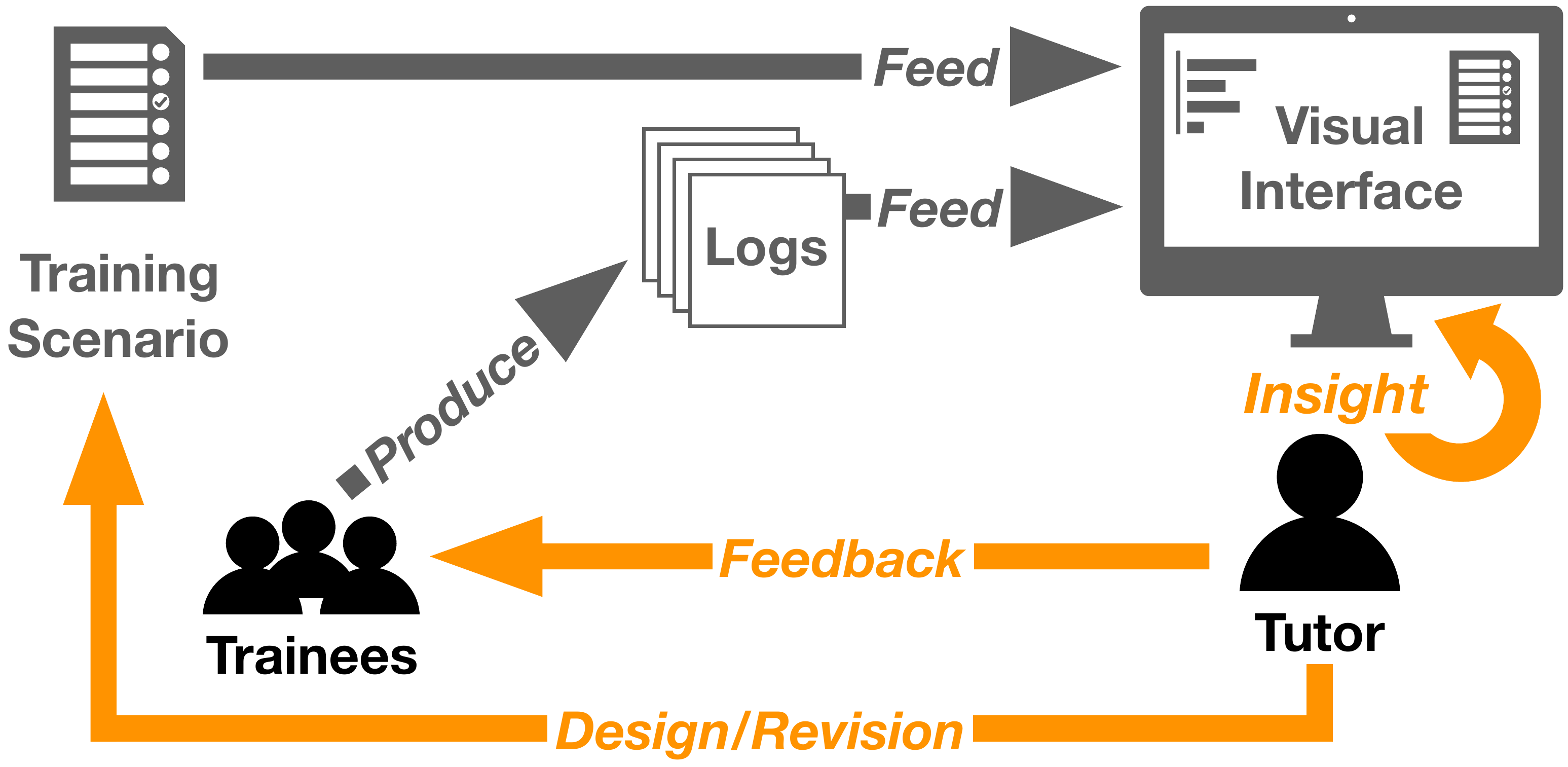}
  \caption{The generalized training workflow. The tutor uses the visual interface to get insight into the training session (to help trainees in trouble) and to revise and improve the training scenario. The data sources are activity logs of the trainees and training scenario description which provides context.}
  \label{fig:workflow}
\end{figure}

Figure~\ref{fig:workflow} illustrates the principal elements and actions of the whole workflow.

There are two broad use cases for the post-training analysis: (a) a comparison of trainees and (b) training scenario improvements. The former is essential when the CTF games are part of the competitions or exams. The rank or grade is then based on the final score and time. However, the tutor cannot understand the subtle difference in the trainee's behavior or expose cheating. Likewise, \emph{training scenario improvements} were usually based on error-prone manual processing of the logged data and anecdotal evidence from training sessions, making revisions inefficient. 

\subsection{Data description}

Hands-on CTF games provide two datasets available for visual analysis: a \emph{training scenario} and timestamped \emph{trainees' events} recorded during the training session. The \emph{KYPO CRP} provides REST API to access these data on-demand in JSON format.

The \textbf{training scenario} contains attributes related to the content. Namely, a background story, puzzle assignments, hints, hint penalties, solutions, solution penalties, correct flags, flag score points, and level time limits. These attributes do not change during the training session. However, tutors might edit them afterward based on trainees' feedback or outcomes from training session analysis. Typical changes include fixing typos and improving the clarity of puzzle assignments, or adjusting level duration estimate, score, and penalty points. 

The \textbf{trainees' events} are automatically collected when trainees interact with the web portal. Example events are: training started, training ended, level started, level ended, correct flag entered, incorrect flag entered, hint taken, solution taken. Each event contains a standard set of attributes (timestamp, event type, training description ID, training session ID, user ID). Three event types (an incorrect flag entered, a hint used, a solution displayed) contain specific attributes -- an incorrect flag string and penalty points.

Although the input data is domain-specific, we can find similarities also in other forms of puzzle-based gaming. Data types are either integers (score and penalty points, level duration estimate -- representing minutes) or text strings (plain-text for flags, markdown markup for all the rest).

\section{Process and methods}
\label{s:process-and-methods}

We closely collaborate with domain experts (cybersecurity educators) from our university who represent target users. They provided initial requirements, gave us feedback on proposed designs, and participated in both evaluations. Our goal was to improve the workflow of tutors and organizers of hands-on cybersecurity training sessions through the design and deployment of the Training Analysis Tool that processes data from the \emph{KYPO Cyber Range}.

In this project, we applied the user-centered approach guided by the design study methodology framework~\cite{Sedlmair2012}, reflecting its \emph{core} stages: discover, design, implement, deploy. Our iterative process has four phases. Each phase reflects one or more of these stages: 

\textbf{Problem characterization} \emph{(discover)}: We conducted semi-structured interviews with three domain experts from the university cybersecurity team. All of them partake in educational activities as seminar tutors or lecturers, and they also participated later on in the evaluation. Each interview lasted about an hour. We also did four field observations during training sessions to gather user requirements and complement our notes, each lasting up to two hours. From these data, we elicited functional requirements and design decisions for both tools.

\textbf{Early prototype and formative evaluation} \emph{(design, implement, deploy)}: We created the early prototype and performed a qualitative formative evaluation with five collaborating cybersecurity educators and one student familiar with the CTF games. 

\textbf{Late prototype and summative evaluation} \emph{(design, implement, deploy)}: We added new features and redesigned the user interface based on received feedback. A qualitative summative evaluation with eight participants served us for the validation of the final designs. 

\textbf{Final deployment} \emph{(implement, deploy)}: The last phase includes the integration of TAT into the \emph{KYPO CRP}. We also plan to collect further feedback from its routine usage. Unfortunately, due to the COVID-19 pandemic, the number of training sessions has been severely limited.

\section{User requirements} 
\label{s:requirements}

Post-training session evaluation provides many opportunities for tutors to perform a detailed analysis of a training scenario and assessment of the trainees.
The interviews and field observations revealed that tutors struggle with analyzing the training data from the individual sessions. They expressed the need for an overview of the data collected during the training session, which enables them to: analyze trainees' behavior, compare their performance, and revise the content and configuration of the training scenario. 

We organized the requirements into the four main categories:

\textbf{R1 -- Trainee behavior analysis:} Tutors should examine trainees' behavior and identify outliers -- e.g., those who are extremely slow/fast or gave up the training. They should assess the trainees by comparing their results (e.g., final time and score, taken hints, number of entered incorrect flags). It is also relevant when the training session is a part of some competition. Further, reviewing the trainees' actions, such as many partially correct flags submitted by several trainees, can point out flaws in the puzzle assignment. 

\textbf{R2 -- Assessment revision:} Correctly set scores and penalties are crucial for the gameplay and trainees' motivation to complete the training. Setting the penalties for hints too small, for instance, can demotivate trainees in attempting to find the solution by themselves. Instead, they could take all hints immediately, which would even result in a better final score. Therefore, the tutors should be able to review the assessment criteria of the training session.

\textbf{R3 -- Timing revision:} Proper estimation of time requirements for cybersecurity puzzles is tricky. Short time allocated for a challenging puzzle can delay the whole session, put unnecessary pressure on trainees to take hints early, or force tutors to intervene prematurely. During the interviews, even the most experienced tutors admitted that they do not have a proper first estimate of mapping puzzle difficulty to time limits. Therefore, tutors should be able to review the time limits of the training session.

\textbf{R4 -- Training content revision:} Tutors should be able to analyze problematic parts of the training content to improve its quality iteratively. The trouble can be hidden either in individual puzzles (e.g., unclear puzzle assignment, useless hint) or their interconnection (e.g., the unbalanced difficulty of two successive levels).

\section{Early design}
\label{s:early-design}

The main goal of the \emph{Training Analysis Tool (TAT)} is to display data from a single training session in the context of the corresponding training scenario (e.g., puzzle assignments, scoring, timing). The tool is designed as a dashboard combining several linked views. Its design follows principles formulated by Oslejsek et al.~\cite{oslejsek2020tvcg}:
\begin{compactitem}
    \item Analyze the \emph{impact of tutor's supervision:} The tool consists of temporal views of trainees' actions and the score development at various levels of detail. Tutors can analyze the impact of both individual and class-wide interventions by focusing on the time of intervention. 
    \item Analyze \emph{quality of training exercise:} All views display the score and time limits that form the primary assessment criteria and delimit the training session's difficulty. These visual artifacts help tutors to analyze the quality of training. Moreover, predefined parameters (penalties, time limits, tasks) are available in the dashboard together with run-time data, enabling tutors to reveal possible weaknesses in training scenarios by comparing expected versus actual development.
    \item Analyze \emph{behavior analysis of trainees:} The training session is captured from several perspectives: temporal view on trainees' activities, a static preview of final results, and detailed dynamic score development. By combining these coordinated views, tutors can interactively analyze individual trainees' behavior, compare them mutually or concerning expected behavior, and visually identify outliers.
\end{compactitem}

\begin{figure*}[!htbp]
  \centering
  \includegraphics[width=.98\textwidth]{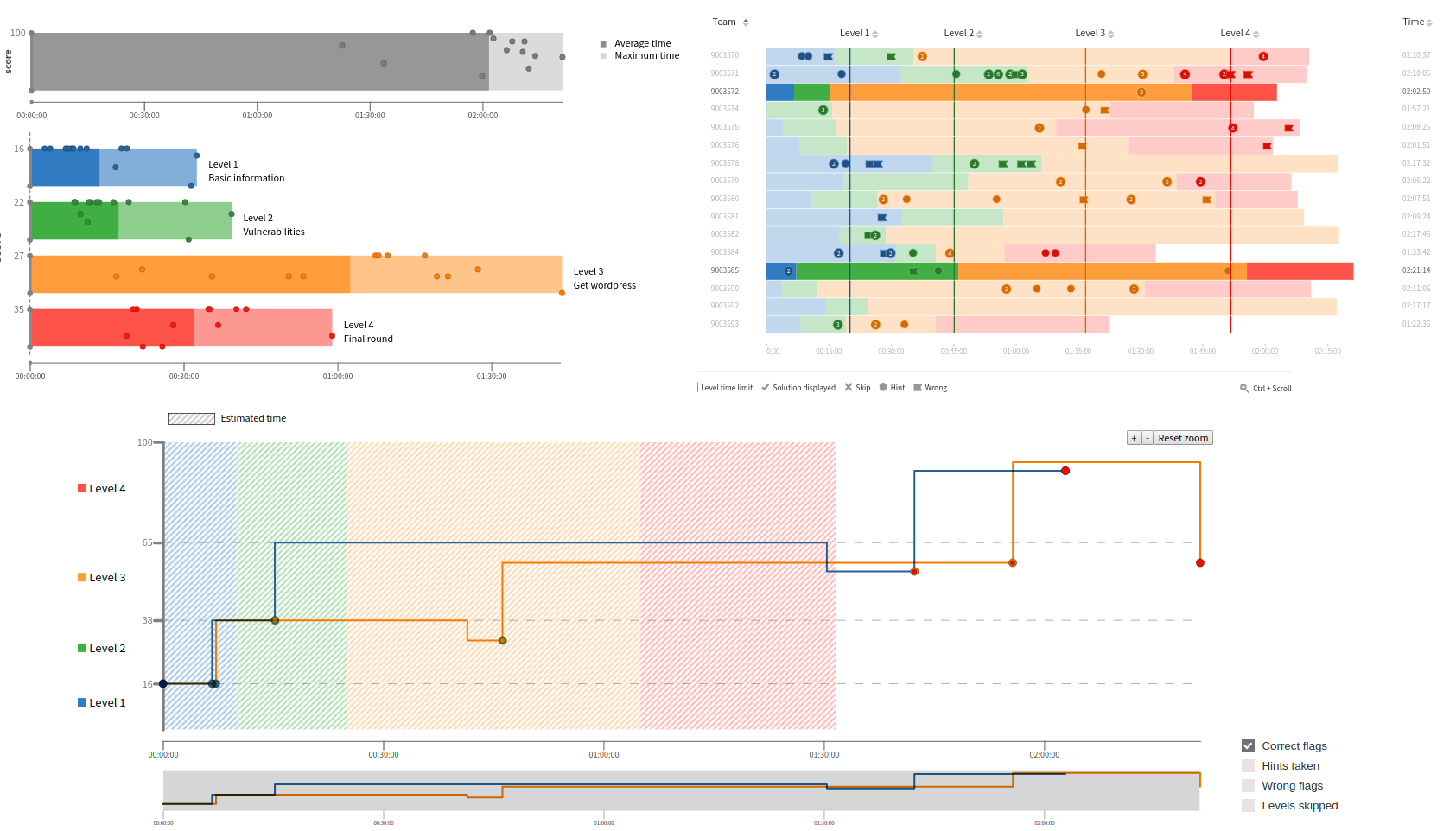}
  \caption{The early prototype of the Training Analysis Tool (TAT) consists of three interconnected visualizations. The \textsc{time-score overview} (top-left) presents the distribution of achieved scores (final and per-level) for each trainee. The \textsc{training overview} (top-right) displays the overall training duration for each trainee and their activities (e.g., taking hints, inserting incorrect flags). The \textsc{individual walkthrough} (bottom) is suitable for a detailed comparison of two or more trainees.}
  \label{fig:early-tat}
\end{figure*}

The early prototype of the \emph{Training Analysis Tool (TAT)} (Fig.~\ref{fig:early-tat}) is a web application consisting of three interactive visualizations: \textsc{time-score overview}, \textsc{training overview}, and \textsc{individual training walkthrough}. 

The former two are based on visualizations proposed by~\cite{oslejsek2019} for player-centered reflection and CTF game results. Since their input data is similar (timestamped events), we used its core design principles and visual encoding, but our visualizations provide extended interaction capabilities. We further elaborate on the design of individual TAT components in detail. 

All three visualizations of the early prototype use a fixed color scheme. The colors were meant to distinguish individual levels of training and were selected in different intensities to be distinguishable for people with the most common forms of color vision deficiencies.

\subsection{Time-score overview}

Total duration and the final score are two main factors used for measuring the performance of the trainees. The \textsc{time-score overview} (Fig.~\ref{fig:early-tat}, top-left) helps identify the correlations between these factors, providing a view on the score distribution, pinpoint the outliers, or allocate clusters. 

Using simple standard statistical views, such as boxplots, would be inconvenient because we need to put in the context multiple metrics (average, and estimate times, final scores). Therefore, the visualization combines bar charts with scatter plots to incorporate time and score data into a single view. The top bar shows the total time (\emph{x}-axis) and each trainee's final score (\emph{y}-axis). The smaller bars below represent individual levels (i.e., tasks). 
Each bar's length expresses the maximum time for the given level (i.e., the time of the slowest trainee). The average time is on the border of two color shades. Although the scoring span can differ in each level, the bars have fixed heights. The vertical space is sufficient to display and analyze achieved score distribution regardless of the scoring span. The maximal level or game score is on the \emph{y}-axis, and the exact score numbers are provided on-demand as tooltips of individual dots together with a trainee's name.

Hovering the mouse cursor over the dot highlights the corresponding results of the trainee in the remaining levels highlight and the exact time and the achieved score for the level display. A mouse click on the dot highlights the corresponding data in the \textsc{training overview} and displays detailed score development in the individual training path at the bottom. Dot clusters can visually indicate the correlations between time and score, which is particularly helpful when the tutor aims to identify the training design issues such as a level difficulty compared to its duration.

The tutors can use it to analyze the results of individual trainees and put them in the context of the training group (\textbf{R1}) or to review score-based assessment (\textbf{R2}). Bar charts also help the tutors review time requirements (\textbf{R3}). Dot clusters may help in the identification of problematic levels in the training scenario (\textbf{R4}).

\subsection{Training overview}

The \textsc{training overview} (Fig.~\ref{fig:early-tat}, top-right) provides a detailed yet compact and uncluttered view of the trainees' progressions and activities. It is based on a stacked bar chart where each row corresponds to one trainee. Segments represent training levels and encompass related game events as glyphs. A user can filter the data based on the level duration and zoom the view to unfold the aggregated events (numbered circles) performed quickly.

The visualization shows the relative time of the training. The stacked bars are aligned to the left, so it is possible to compare the time requirements regardless of the delays caused by individual trainees' various starting times (\textbf{R3}). Level labels above the bars support sorting by the duration of the corresponding levels. The related vertical lines indicate the expected level duration. When sorted, they also reveal the deviation of the actual and estimated time for each trainee.

The glyphs indicate events. In this view, they help the tutors to recognize possible problems in the design of training definition (\textbf{R4}) or analyze the behavior of the trainees (\textbf{R1}). For example, multiple incorrect flags submitted by diverse trainees can indicate unclear or ambiguous instructions; many hints taken in quick succession may suggest a lack of effort caused by improper difficulty. 

\subsection{Individual walkthrough}

The \textsc{individual walkthrough} (Fig.~\ref{fig:early-tat}, bottom) is based on a step chart with glyphs representing trainees' actions. It enables the tutors to track outliers' behavior (\textbf{R1}) and explore the cause of recognized problems in the training session (\textbf{R2}, \textbf{R4}). It provides a detailed insight into a trainee's advancement and actions or allows comparing two or three trainees selected from the \textsc{training overview} list or the \textsc{time-score overview}. The y-axis represents gained score. The horizontal dashed lines imply the maximal level score. The striped background outlines the estimated level times. 

A zoom function allows adjusting the view on a selected portion of the chart, which is useful when the events are clustered.
On mouse hover, a tooltip shows details for each action. A context view frame below the main chart helps the tutor to get oriented in the zoomed area and shift the time range when needed. Furthermore, the checkboxes in the bottom right corner allow filtering the event types.

\section{Formative evaluation}
\label{s:formative-evaluation}

The main goal was to gain feedback on the TAT's usefulness in four areas:
\begin{compactitem}
    \item {\bf Trainees} -- Is it possible to identify trainees who struggled (e.g., lacking behind, stuck with the task/level)? Can tutors recognize any unusual behavior of trainees (e.g., cheating, prolonged inactivity)?
    \item {\bf Training session} -- Is it possible to recognize when the training is running out of schedule? Can tutors identify scenario design issues?
    \item {\bf Visual encoding} -- Is the visualization easy to understand? What type of information is redundant or missing? 
    \item {\bf Interaction} -- How do tutors interact with the visualization? Are the interaction capabilities sufficient?
\end{compactitem}

We further evaluated the usability and usefulness of the visualizations and gathered remarks on visualization improvements for the following design process iteration.

\subsection{Participants}

Due to the necessary background knowledge of hands-on cybersecurity training, we conducted a qualitative user study with five domain experts (P1--P5) and one student (P6). All of them were members of the university cybersecurity team who partake in hands-on training on different positions. Table~\ref{tab:demography} shows their demographic information. 

\begin{table}[!htpb]
    \footnotesize
    \caption{Demographic summary of the participants and their involvement in the design study. TE -- teaching experience (in years), OE -- organized hands-on exercises (in sessions). Participation in individual stages: PC -- problem characterization; FE -- formative evaluation; SE -- summative evaluation.}
    \label{tab:demography}
    \centering
    \begin{tabular}{cclccccc}
        \textbf{ID} & \textbf{Age} & \textbf{Position} & \textbf{TE} 
        & \textbf{OE} & \textbf{PC} & \textbf{FE} & \textbf{SE}\\
         \\
         P1 & 33 & Lecturer, Manager & 4  & $>$20 & \checkmark & \checkmark & \checkmark \\
         P2 & 27 & Seminar tutor     & 7  & $<$20 & \checkmark & \checkmark & \checkmark  \\
         P3 & 31 & Seminar tutor     & 3  & $>$20 & \checkmark & \checkmark & \checkmark  \\
         P4 & 27 & Seminar tutor     & 5  & $<$10 & & \checkmark & \checkmark \\
         P5 & 35 & Senior lecturer   & 5  & $>$20 & & \checkmark & \checkmark  \\
         P6 & 24 & CTF Course graduate  & 0  & 1 & & \checkmark & \checkmark \\
         P7 & 22 & CTF Course graduate  & 0  & 1 & & & \checkmark \\
         P8 & 21 & CTF Course graduate  & 0  & 0 & & & \checkmark \\
    \end{tabular}
\end{table}

\subsection{Procedure}
\label{s:se-procedure}

In September 2019, we held the formative evaluation sessions in person using 27" iMac with the resolution 2560$\times$1440 and Google Chrome browser version 76. The experimenter took notes and audio recorded the participants' opinions and thoughts. 

The user study had two parts, and the participants were asked to think aloud. The sessions lasted about an hour. In the first part, the experimenter outlined the procedure. The participant consented and filled the demography questionnaire. The experimenter presented the TAT and situated the participant in the role of a tutor using the tool. Next, the participant spent 2--3 minutes familiarizing with it using dummy data followed by completing three tasks addressing requirements R1--R4: 
\begin{compactitem}
     \item \emph{T1: Identify an unusual behavior of trainees and name the potential issues.}
    \item \emph{T2: Find and compare a pair of trainees who: a) have the same score; b) were the best and the worst; c) were the slowest and the fastest. How do they differ?}
    \item \emph{T3: Identify problems caused by the poor design of the training scenario and propose improvements.}
\end{compactitem} 

Participants performed the tasks on two data sets DS1 and DS2. We chose the genuine data since they contain various actions observable during training sessions (e.g., guessing the correct flag, prolonged inactivity, varying trainees' performance). Their different size, number of trainees, and duration show two distinct yet ordinary real-world circumstances.

DS1 is from the tutorial on computer forensic skills and consists of six game levels. The goal is to identify and examine malicious software running in the computer system. The trainees learn how to identify a suspicious application, dissect its executable, and process memory. The session lasted 55 minutes, and 16 trainees generated 374 events, making the 23.4 events per trainee on average. DS2 is an attack-oriented training scenario that consists of four game levels with the following puzzles: exploit server vulnerability, gain the root privileges, access a protected data file, and cover the traces after the attack. Six trainees generated 146 events over 90 minutes, averaging 24.8 events per trainee.

Finally, the participant filled two usability questionnaires and was debriefed.
We chose the SUS -- System Usability Scale~\cite{Sauro2011} and the SEQ -- Single Ease Question~\cite{Sauro2009}, two widely used questionnaires for measuring various products' usability. The former is a widely used method for assessing the usability of the systems. The latter is considered a robust measure to quantify the usability for tasks that are too complex for metrics like task duration time or completion rate\footnote{The user responds to a single precisely-worded question (``Overall, how difficult or easy did you find this task?''), using a scale from 1 (Very difficult) to 7 (Very easy).} and when the number of participants is low, as in our case.

\subsection{Results}
\label{s:results}

The formative evaluation revealed weaknesses in the early design and helped us understand tutors' work after the training session. 

The most acclaimed feature of the \textsc{training overview} visualization is the ability to sort trainees by the time spent at some level and compare them to the estimated level duration (defined in the training scenario). Participants also used the visualization to identify the trainees who significantly exceeded the estimated level duration time.

For most of the participants, the \textsc{score overview} visualization was a starting point when solving all the tasks. They used it to identify outlying trainees (P2, P4, P6), to assess the difficulty of each level based on the time/score distribution of trainees (P1, P2, P4, P6), or to compare it with the maximum score per level (P3, P4). P3 also used the score overview to assess the conceptual design of the training scenarios (the first levels should be manageable and short compared to the final ones).
Participants lacked information about estimated level duration (P1--P4, P6). P6 wanted even more details, such as medians of time and score for each level.

Participants often used score overview visualization to highlight trainees in training overview and vice versa. Score overview was also often used in T2 as a selector for trainees to compare. We did not observe any other extensive mutual use of two or all three visualizations. On the other hand, the \textsc{individual training walkthrough} visualization was generally considered "useful only in a specific case when the training session is organized as a competition to decide the final order of trainees" (P4). 

The main complaint (mentioned by all) was the absence of a tabular view showing various details of all trainees such as their final score, scores per level, number of taken hints, or incorrect flags.

Other frequent issues were: the absence of filtering features (P1--P5); a missing overview of the training scenario allowing the users to skim through the texts of tasks, set penalties, and flags (P2--P4, P6); insufficient integration of the visualizations (P1, P2, P4, P5); and the visual encoding (P1, P3, P4, P6) considered by P3 as "disturbing due to many colors without proper meaning."

\begin{figure}[!htbp]
  \centering
  \includegraphics[width=.48\textwidth]{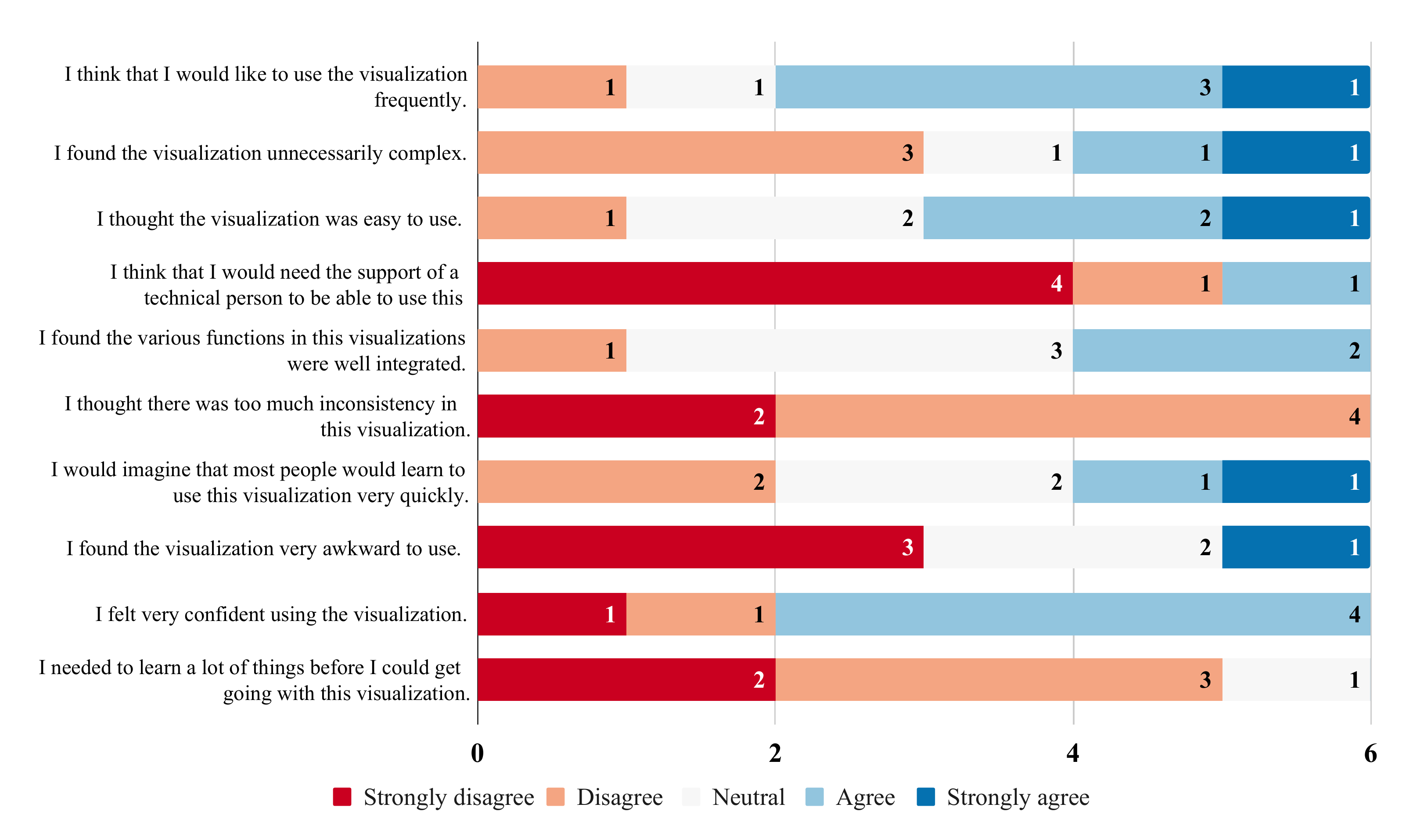}
  \caption{Formative evaluation: The SUS questionnaire responses.}
  \label{fig:fe-sus}
\end{figure}

The SUS score was 65.4 points (out of 100). It corresponds to the \emph{good} rating, according to the adjective ratings~\cite{Bangor2009}. Fig.~\ref{fig:fe-sus} summarizes the SUS questionnaire responses. With the SEQ score of 6.5 (out of 7), the TAT showed to be well-suited for training design analysis (T3: Identify training design issues.). The two tasks focused on identifying and comparing trainees scored 5 (T1: Unusual behavior of trainees) and 6 (T2: Comparison of trainees). 

While these results confirmed the overall usability and usefulness of the TAT, we had to address the main issues raised by the study participants.

\begin{figure*}[!htb]
  \centering
  \includegraphics[width=\textwidth]{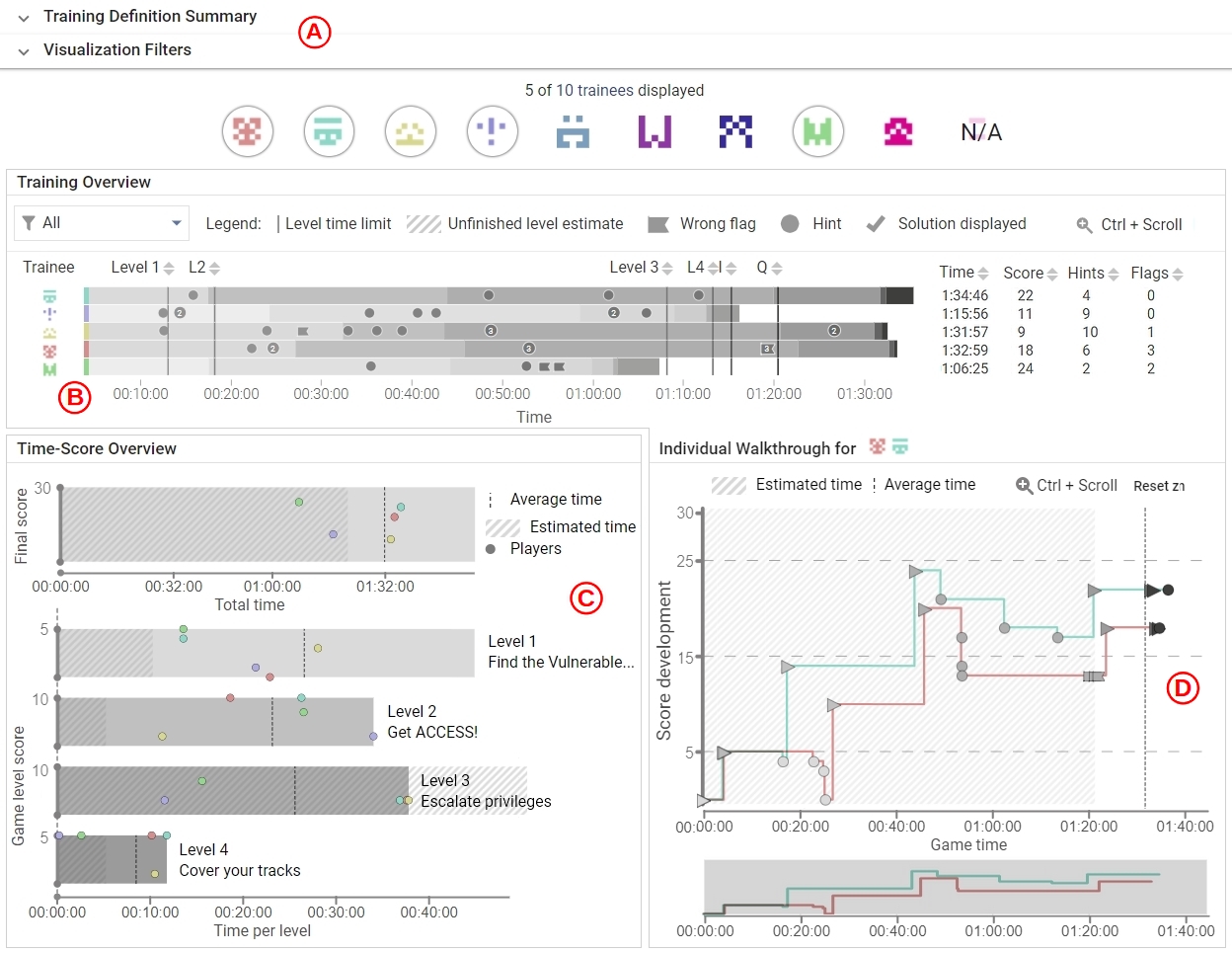}
  \caption{The Training Analysis Tool (TAT) consists of the upper panel for training definition summary and filters (A) and three visualizations: the \textsc{training overview} (B) displays the overall training duration for each trainee and their activities (e.g., taking hints, inserting incorrect flags). The \textsc{time-score overview} (C) presents the distribution of achieved scores (final and per-level) for each trainee. The \textsc{individual walkthrough} (D) directly compares of two or three trainees and is subordinate to the \textsc{training overview}.}
  \label{fig:tat}
\end{figure*}

\section{Final design}
\label{s:design}

We revised the final design rationale, visual encoding, and interaction capabilities of the current version of the TAT based on the formative evaluation. The prototype, implemented using Angular and D3.js library, is available at \url{https://tat.surge.sh}. 

The main principles of the three visualizations remain the same. However, we significantly redesigned the layout making the \textsc{training overview} the most prominent visualization. We also added more filtering options for selecting individual trainees and revised the use of colors. The formative evaluation also revealed that the coloring of levels is not essential for the users, so we have changed it in the late prototype: the platform on which the training sessions take place generates a unique avatar for each trainee. Therefore, we decided to emphasize the trainees based on the avatar's color 
instead. Now, each trainee has a unique color in all three visualizations. These colors are not intended as the exclusive means of trainee identification but as complementary visual support (to accompany the ability to highlight or filter the trainees). To distinguish training levels, we used gray color shades in the late prototype.

Finally, we added additional information regarding the training definition, such as the task descriptions, correct flags, and contextualized trainees' data with individual levels. Fig.~\ref{fig:tat-detail} displays the final layout, with the collapsed \textsc{training definition summary} and \textsc{visualization filters} sections.

\subsection{Training definition summary and visualization filters}

\begin{figure}[!h]
 \centering
 \includegraphics[width=.48\textwidth]{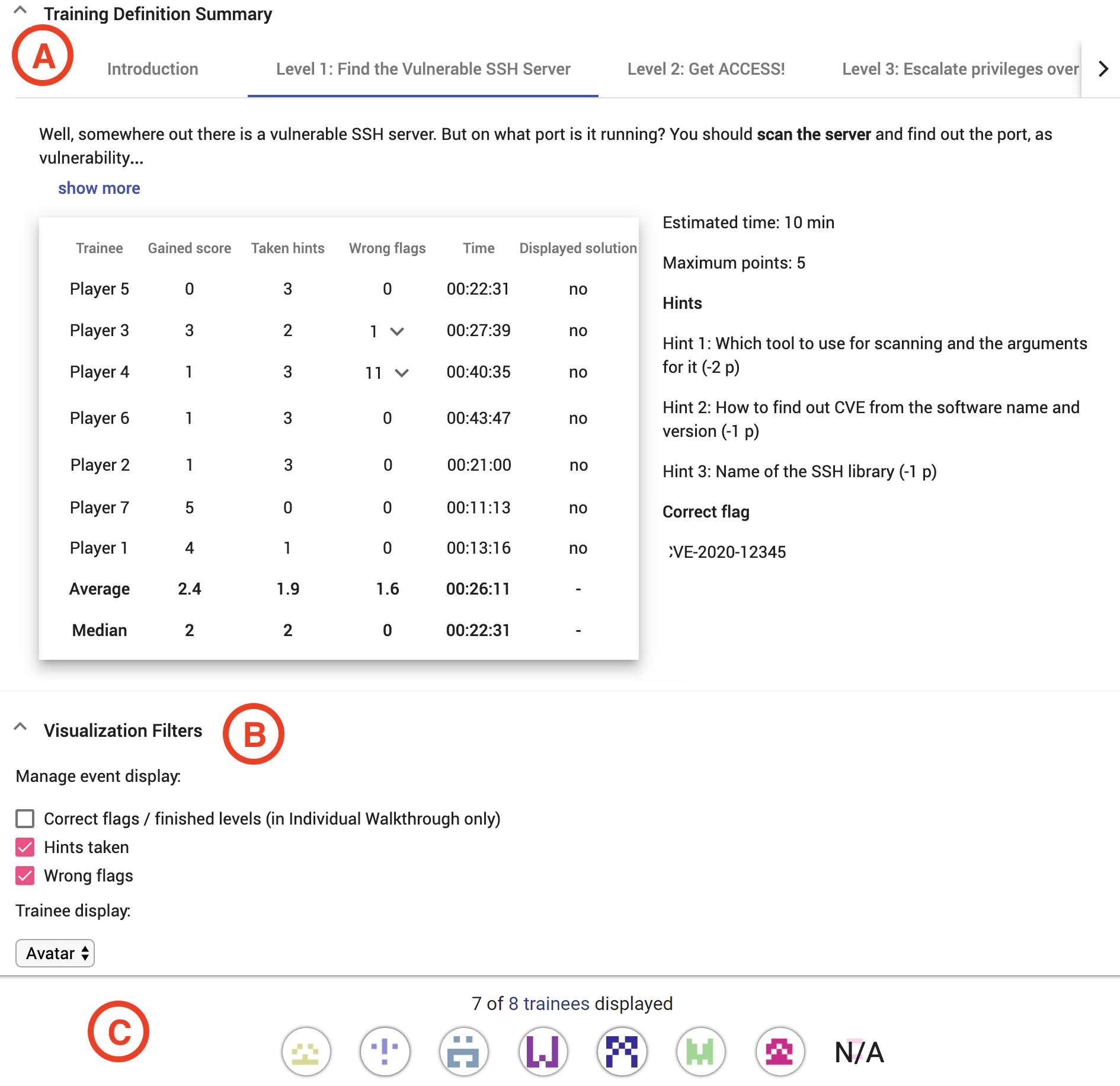}
 \caption{TAT -- details of the \textsc{training definition summary} (A), \textsc{configuration} (B), and the \textsc{trainees} (C) sections.}
 \label{fig:tat-detail}
\end{figure}

The TAT's upper part (Fig.~\ref{fig:tat} -- A) contains a collapsible panel with the training definition details, visualization filters, and avatar-based trainees filter. The \textsc{training definition summary} serves for the configuration of the tool and synopsis of the training. It provides training scenario parameters (i.e., task assignments, hints, penalties, correct flags). The tabs show data for individual levels (Fig.~\ref{fig:tat-detail} -- A). For each game level, a table summarizing data of individual trainees provides an overview of the gained score, taken hints, incorrect flags, and time spent in the level (\textbf{R2} and \textbf{R3}). Comparing the results shown in the table with the level content and parameters (e.g., the comparison of incorrect flags with the correct flag or scheduled time allocation with the average or median values) can help the tutors identify problematic parts of the content (\textbf{R4}). 

The \textsc{Visualization Filters} (Fig.~\ref{fig:tat-detail} -- B) are global filtering options to show or hide glyphs representing hints or flags and switch between trainees' avatars and names (IDs). The avatars (Fig.~\ref{fig:tat-detail} -- C) are switches for filtering out the trainees from the \textsc{training overview} and \textsc{time-score overview}. 

\subsection{Training overview}

We extended the \textsc{training overview} (Fig.~\ref{fig:tat} -- B) with the table summarizing total game duration, achieved score, number of taken hints, and submitted incorrect flags for each trainee.
We also added the legend for quicker orientation.

The \textsc{training overview} interacts with two complementary views. By clicking on the stacked bar, the \textsc{individual walkthrough} visualization appears, showing score polyline and events of the corresponding trainee. The level bars highlight the corresponding dots in the \textsc{time-score overview} and the polyline in the \textsc{individual walkthrough} on mouseover. 

\subsubsection{Time-score overview}

\begin{figure}[!h]
  \centering
  \includegraphics[width=.35\textwidth]{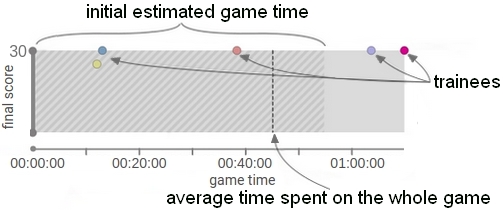}
  \caption{The \textsc{Time-Score Overview} combines bar charts with scatter plots to show relationships between the score and time of the game levels.}
  \label{fig:tso}
\end{figure}

Unlike the early prototype version, we added the dashed vertical line to indicate the actual average completion time of the trainees. The striped segments delimit the time estimate for each level. Therefore, the tutors can quickly identify the differences between the expected and the actual (and averaged) time for each level, as shown in Fig.~\ref{fig:tso}.

\subsubsection{Individual walkthrough}

In the final version, the \textsc{individual walkthrough} (Fig.~\ref{fig:tat} -- D) displays upon selecting a trainee in the \textsc{training overview}. The selected trainees are indicated as avatars next to the title.
We also reflected the main complaints regarding the clutteredness and simplified the visualization layout. Only the total training duration estimate is shown instead of the estimate for each level. We also added the vertical dashed line to indicate the actual average time, similarly to the \textsc{time-score overview}.

\section{Summative evaluation}
\label{s:evaluation}

The summative evaluation was held in April 2020. We intended to validate the final design concerning the user requirements \textbf{R1--R4}, assess the usability and usefulness of the TAT, and identify possible refinements for the final integration into the \emph{KYPO CRP}.

\subsection{Participants}

We asked the same six people who participated in the formative evaluation. We also recruited two more students who passed the CTF design course taught at our university (see Table~\ref{tab:demography}). They represent novice users familiar with CTF games' basic concepts and only have hands-on experience with their design.

\subsection{Procedure}

Due to the COVID-19 pandemic restrictions, we held it remotely using Google Meet, which we also used to record audio and screen. The participants used their computers or laptops with the 13.3"--27" screens and resolutions ranging from FullHD to UHD. The procedure was almost the same as in the formative evaluation (see Sec.~\ref{s:se-procedure}). The only difference was a new data set that we used for the tasks.

DS3 uses data from a training session held as the introductory lecture of the CTF game design course of Fall 2019. It is an attack-oriented four-level training scenario similar to DS2, in this case, tested on nine trainees who generated 281 events in the session lasting 110 minutes. On average, each participant performed 31.3 events.

\subsection{Results}

The participants completed all the tasks without struggle. Despite minor difficulties, the immediate feedback was more positive than in the previous evaluation. Since the tasks are complex and depend on the tutor's knowledge and experience we sought qualitative input rather than measuring user performance.

Participants mostly worked with the \textsc{training overview} since it contains most of the necessary information. The \textsc{time-score overview} serves well to identify timing issues and assess level difficulty. The \textsc{training definition summary} supports finding flaws in the puzzle assignments (e.g., misleading texts, wrong instructions for flag format). Further, we did an inductive qualitative analysis~\cite{Thomas2006} of the video recordings, which is summarized below.

\begin{figure}[!htbp]
  \centering
  \includegraphics[width=.2\textwidth]{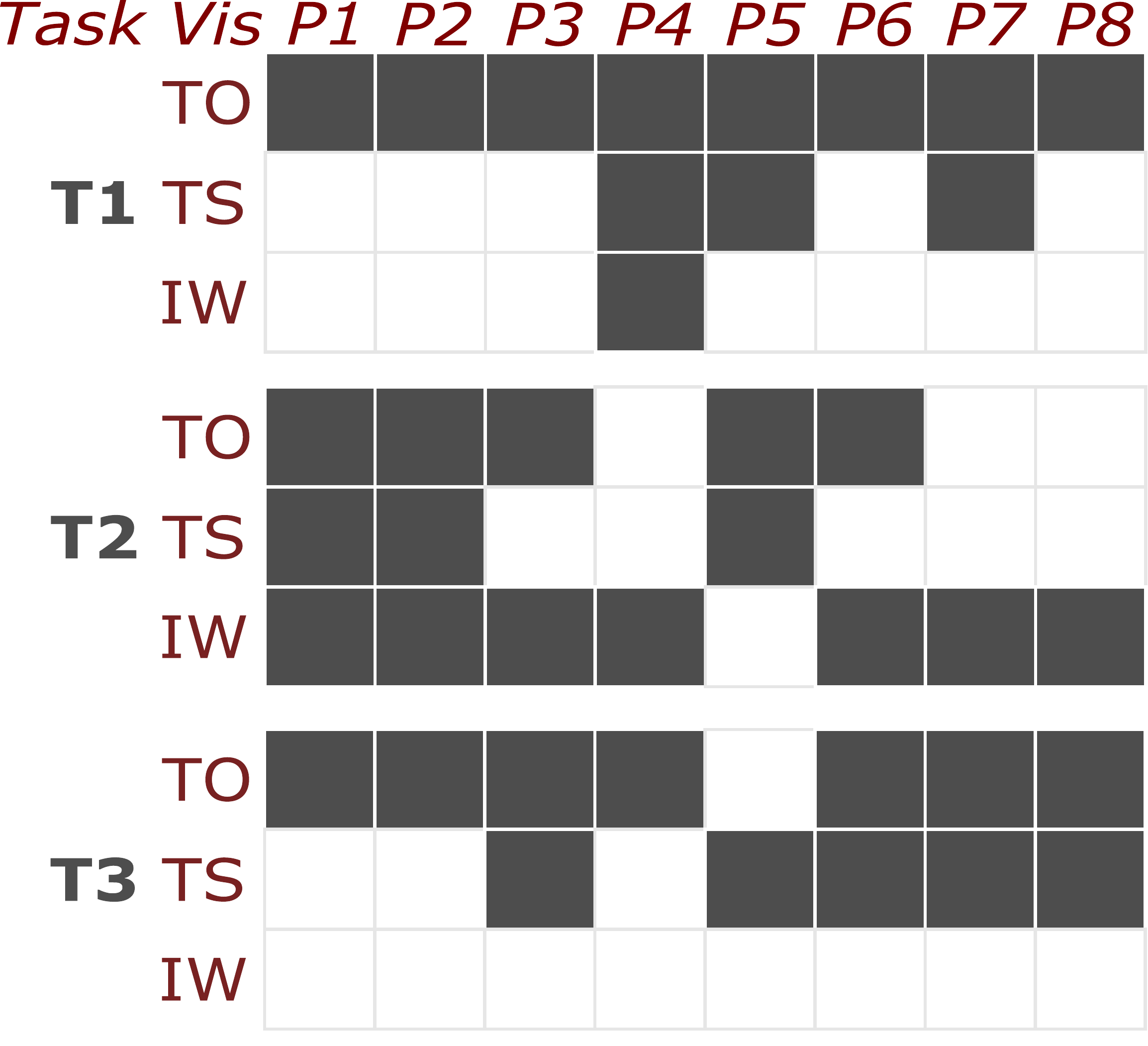}
  \caption{Gray cells indicate visualization usage (Vis) when solving tasks T1--T3 for each participant (P1--P8). Visualizations: Training Overview (TO), Time-Score Overview (TS), Individual Walkthrough (IW).}
  \label{fig:visualization-usage}
\end{figure}

\textbf{Visualizations usage.}
Figure~\ref{fig:visualization-usage} shows the usage of visualizations to solve the tasks by participants. The most preferred was the \textsc{training overview}. All but P5 used the \textsc{training overview} as a starting point when solving all the tasks (P5 preferred the \textsc{time-score overview}). Its most acclaimed feature is the ability to sort trainees by the time spent in individual levels and compare them to the estimated level duration (defined in the training scenario). Participants also used the visualization to identify the trainees who significantly exceeded the estimated level duration time. All the sorting options (by time spent in a level, final time, score, hints, and incorrect flags) were used at least once by each participant.
On the other hand, the zooming function was used only rarely (P1, P6). The participants used the \textsc{time-score overview} to identify outlying trainees (P2, P4, P5), assess each level's difficulty based on their time/score distribution (P1, P2, P4, P5), or compare it with the maximum score per level (P3, P4). The \textsc{individual walkthrough} was still considered the least usable (P1, P2, P5, P6, P7). P1 and P5 did not work with it at all. Others used it only for a direct comparison of two trainees (\emph{T2}). 

\textbf{The TAT allows comparison of trainees beyond time and score.}
To identify non-standard trainees' behavior (\emph{T1}), we observed that all the participants revealed all or almost all occurrences of the most common types, such as taking all hints at once shortly after they entered a new level or guessing the flags in each dataset.
The participants found those with the lowest score/largest time, followed by a detailed inspection of the number of taken hints and inserted incorrect flags. The procedure was the same for all. The difference was only in the starting visualization. While P4, P5, and P7 started with the \textsc{time-score overview}, the rest used the \textsc{training overview} solely. The participants also intensively used the trainee filter combined with the \textsc{training overview} sorting capabilities to filter out unwanted trainees quickly, especially for the second task (\emph{T2}). Despite the \textsc{individual walkthrough} received mixed reactions, most participants (except P1 and P5) used it for a head-to-head comparison. 

\textbf{The TAT helps to identify training scenario shortcomings.}
When dealing with the identification of training scenario shortcomings (\emph{T3}), the participants mainly focused on three areas: correcting the time estimates and maximal score of individual levels, the perceived level difficulty, and instructions for a correct flag format. All the participants proposed changing the time estimates or the assigned maximum of points based on the trainees' overdue in the first two levels of D3. Moreover, seniors (P1, P3--P5) also identified the confusion with the flag formatting instructions in the second level. P3--P5 analyzed the data even more profoundly and revealed the flaw in the game design based on the observation that some trainees used the correct flag for the fourth level in the third one.

Except for P1, P2, and P4, the participants used the \textsc{training definition summary} since it clearly shows the difference between the estimate and real-time. The size of each level allows for a quick comparison of their perceived difficulty (the longer it took, the problematic the level was). The glyphs visualizing incorrect flags in the \textsc{training overview} proved to be good indicators for potential issues with the puzzle assignments, including the technical instructions. All the experts (P1--P5) greeted the \textsc{training definition summary} as a convenient way to search for problematic parts of the training definition. 

\textbf{Gaps and drawbacks of the TAT.} 
We received several suggestions for further improvements to the TAT visualizations. P5 suggested adding ``the horizontal line also showing the average score per level'' in the \textsc{time-score overview} to improve comparing level scores. The two-level filtering (avatars $\rightarrow$ trainees in the \textsc{training overview}) received mixed feedback. Only three participants (P1, P2, P5) used both to filter out specific trainees, while others preferred to keep all of them visible. The evaluation also revealed that with the grayscale for the \textsc{training overview}, highlighting of selected trainees is not very pronounced and will be revised in future development.

The main benefit of the \textsc{individual walkthrough} is that the polyline visualizing score development better informs the tutor whether there are similarities in the trainees' gameplay. Since this is useful only in a specific use case, we will reconsider its integration in the subsequent design iterations simplifying the user interface.

\begin{figure}[!t]
  \centering
  \includegraphics[width=.48\textwidth]{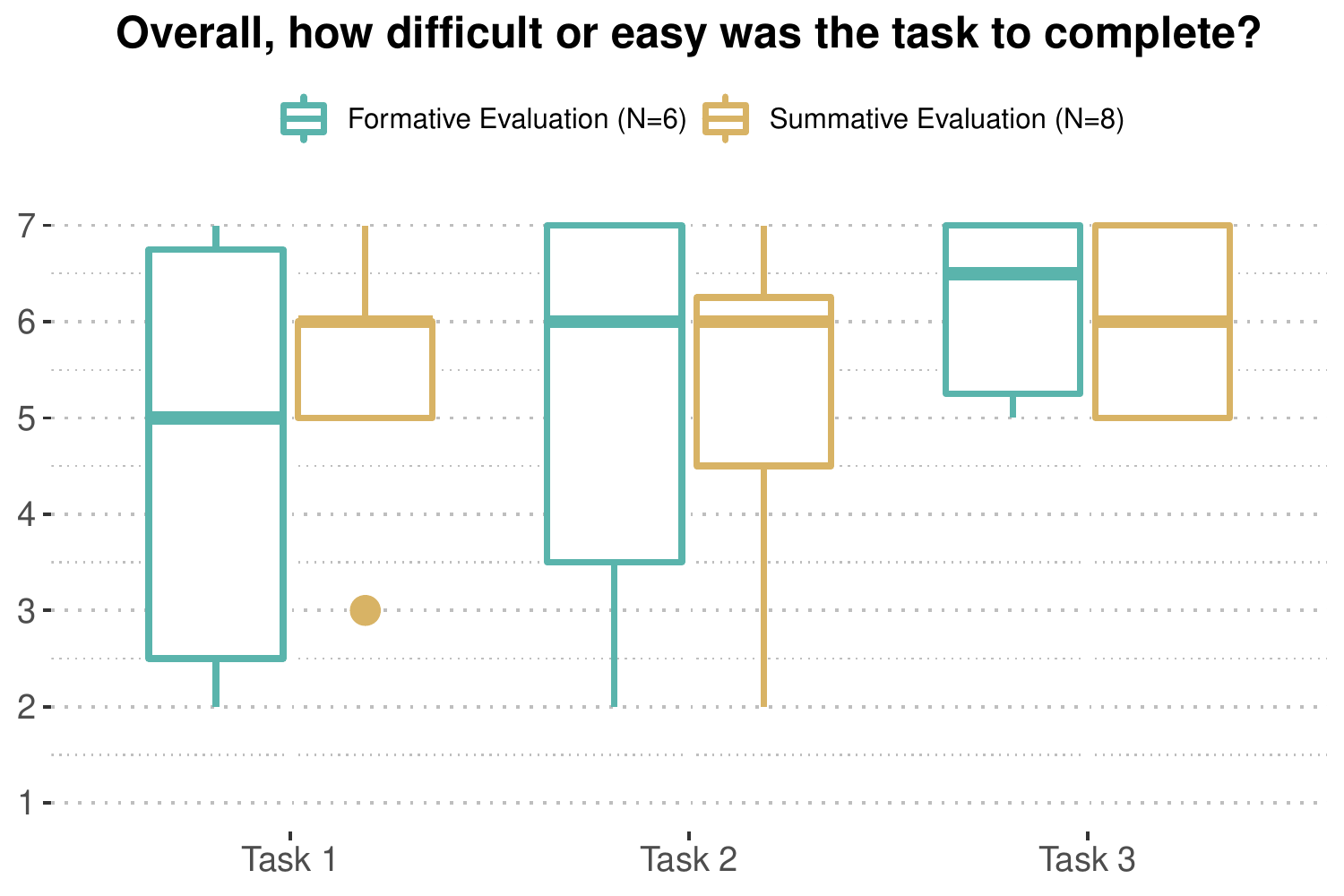}
  \caption{Single Ease Question scores of the tasks in both evaluations.}
  \label{fig:seq}
\end{figure}

\begin{figure}[!t]
  \centering
  \includegraphics[width=.48\textwidth]{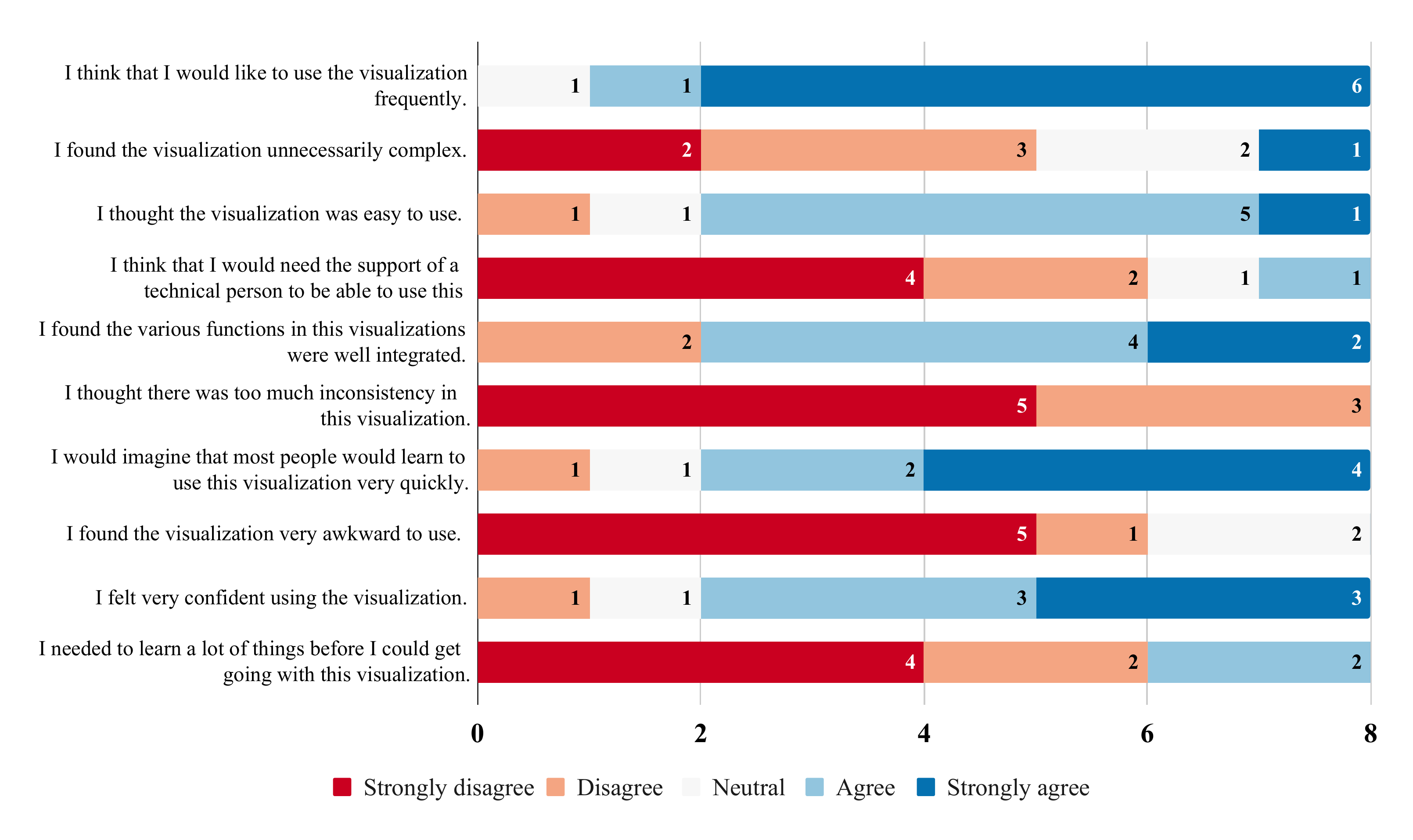}
  \caption{Summative evaluation: The SUS questionnaire responses.}
  \label{fig:se-sus}
\end{figure}

The average SUS score raised to 77.5 (compared to 65.4 for the early prototype), which still equals to \emph{good} rating. We assume that it is mainly due to the higher complexity of the tool and the remaining issues with the \textsc{individual walkthrough}. The data plot of the SUS questionnaire responses is in Fig.~\ref{fig:se-sus}. 
However, the medians 6.0 of SEQ score (Fig.~\ref{fig:seq}) for all the tasks (\emph{T1--T3}) further supports our statement that the TAT is well-suited for the post-training analysis.

\section{Discussion}
\label{s:discussion-and-future-work}

In this section, we discuss the findings and limitations of the studies.

\subsection{Lessons learned}

The summative evaluation validated our design decisions. The verbal feedback from participants and the SEQ and SUS scores confirmed that the tools address the elicited requirements. We also revealed three notable findings regarding the presentation of summaries, sorting and filtering capabilities, and domain specificity.

\textbf{Summaries.} 
Extending the visualization with pertinent summary data could help tutors to overview the situation and identify anomalies quickly. Especially in analytical tools, even elementary statistics and simple charts are helpful. Although we did not implement such charts in the TAT, some participants asked for them as feature requests.

\textbf{Sorting and filtering.} 
The evaluation revealed that we should work with the sorting and filtering options even more thoughtfully so that tutors can better focus their attention. There must be a real usage scenario for each filter type. Particular attention should be paid to carefully selecting items for filtering and the batch selecting and filtering shortcuts (e.g., ``deselect all'').

\textbf{Domain-specific insight over universality and scalability.} Puzzle-based learning represents a vast area where tutors' support tools differ vastly among various application domains. Since there are no guidelines or best practices and the user requirements are often contradictory, they have to be considered carefully, and the tools should be tailored to specific uses.
Furthermore, the amount of data from a single session is usually relatively small.

\subsection{Limitations}

Both user studies had two main limitations to the external validity: low number of participants and qualitative focus of the evaluation in the controlled environment instead of the in-the-wild evaluation.

To ensure the evaluation's ecological validity, we needed users with practical experience with organizing hands-on training sessions and knowledge of cybersecurity education. These demands notably restrict our choice of suitable candidates. Our collaborating cybersecurity educators are, no doubt, the primary users of the developed tools. Therefore, they provided relevant feedback, which will serve as a source for further thoughts on both tools' improvements. We also asked students of the cybersecurity degree program who successfully passed the university course on CTF games design. They represent novice users unfamiliar with analytical visualizations.

Due to the qualitative nature of the evaluations, we did not focus on finding the limits in terms of the total number of trainees and their events since the events with more than 16 participants are literally none due to the space limits of the training facility at our university. We originally planned to perform the case studies to assess the TAT's final design in the actual deployment. Unfortunately, due to the COVID-19 pandemic, the scheduled hands-on training sessions had been canceled, and the only feasible option was to perform the evaluation remotely, using the same procedure as in the summative evaluation. 

In this work, we restrict ourselves to the case study of hands-on cybersecurity courses focused on system hacking and cyber-attacks. In particular, puzzle-based capture the flag games where the structure and data are well-defined in advance. These restrictions allowed us to provide the tutors with a more in-depth insight into this specific application sub-domain through a pair of visualization tools. 

Despite these limitations, the provided feedback has been guiding our work and feature requests for the deployment into the \emph{KYPO CRP}.

\section{Conclusion and future work}
\label{s:conclusions}

We introduced the visual analytics tool that, based on the qualitative feedback, improves the tutors' insight into the training sessions and allows them to assess the quality of the training scenarios and evaluate the training session results. We focused on low-level learning analysis (i.e., analyzing data from a single training session). As we pointed out in Sec.~\ref{s:related-work}, this particular area is often overlooked since the main focus in support tools for tutors and educators is on high-level analysis for MOOC e-learning. 

We have presented a design study on applying visual analytics to data from hands-on cybersecurity training in the form of CTF games.
We introduced two iterations of the \emph{Training Analysis Tool}, allowing tutors to assess the quality of the training scenarios and gain insight into the trainees' progress beyond the completion time and final score. The summative evaluation validated our design decisions. The verbal feedback from participants and the SEQ and SUS scores confirmed that the tools address the elicited requirements.
We gradually learned more about what information tutors would like to display in the visualization and how they interact with the data during the design study. Based on this experience, we believe that a data-driven insight into the training courses could provide surprising insights and knowledge about the design and behavior of trainees.

Focusing on puzzle-games principles enabled us to conceptualize the data and visualizations beyond the cybersecurity domain. If we look closely at the information we used, we realize that it is a quadruple: timestamp, the ID of the trainee, type of event, content (arbitrary). Therefore, we believe that our approach can be easily applied in other areas where hands-on training becomes common. We admit that there are further requirements, such as automated processing of user inputs, but even basic logging can provide sufficient data. The level of detail depends mainly on the expressiveness of the content component.

Consider the university programming course as another application area. The tutors often evaluate students' assignments using automated compilation and validation tools against predefined unit tests and datasets. The summary of code diffs, compiler error logs, and output of the automated tests can be logged. Similar to the cybersecurity domain, these events can be mapped to assessment events (e.g., penalties for unsuccessful unit tests), player actions (e.g., the submission of a piece of code), and progress events (e.g., successful compilation and test of a programming task). Visualizing these events on the timelines (one per student) or further text analysis of the code can be as valuable as our analogy with the cybersecurity CTF games.

The support tools for a category of so-called blended classrooms and hands-on courses are still mostly unexplored. Our work addresses only a tiny part of this broad research area. Despite our focus on cybersecurity education, we consider our findings applicable in other areas of puzzle-based learning and analyzing data from a single training session (i.e., low-level learning analysis). We want to encourage others to explore novel methods for visual analysis of puzzle-based learning courses in different areas.

The TAT is integrated into the user interface of the \emph{KYPO CRP}. We also work on additional data integration from sandboxes (e.g., resource usage, executed commands, running processes). Enhancing the current level of event processing with this information will further improve the insight and enable a more detailed analysis of the training and its scenario.
Our next goal is to explore the possibilities for visual analysis of multiple training sessions and analyze and assess trainees' long-term progress. Extending the analysis with automatic highlighting of anomalies or flaws in the training design is another direction of research that needs further study.

\bibliographystyle{cag-num-names}
\bibliography{refs}

\end{document}